\documentclass[12pt]{article}
\usepackage{times}
\usepackage{geometry}
\usepackage[utf8]{inputenc}
\usepackage{enumitem,amssymb}
\usepackage{ragged2e}
\usepackage{graphicx}
\usepackage{parskip}
\newlist{thematic}{itemize}{8}
\setlist[thematic]{label=$\square$}
\usepackage{pifont}
\newcommand{\cmark}{\ding{51}}%
\newcommand{\done}{\rlap{$\square$}{\raisebox{2pt}{\large\hspace{1pt}\cmark}}%
\hspace{-2.5pt}}

\usepackage{hyperref,xcolor,amsmath}
\usepackage[sort&compress]{natbib} %Bibliography Management
\usepackage{aas_macros} % Translates Journals Shorthand in Bibliography 

\begin{document}
\raggedright
\huge
Astro2020 APC White Paper \linebreak

2020 Vision: \vspace*{10pt}\linebreak
Towards a Sustainable OIR System  
 \linebreak
\normalsize

\noindent \textbf{Thematic Areas:} \hspace*{60pt} $\done$ Ground Based Project \hspace*{10pt} $\square$ Space Based Project \hspace*{20pt}\linebreak
$\done$ Infrastructure Activity \hspace*{31pt} $\done$ Technological Development Activity \linebreak
  $\done$  State of the Profession Consideration \hspace*{1pt} $\square$ Other \hspace*{40pt} \linebreak
 
 % $\square$    Galaxy Evolution   \hspace*{45pt} $\square$ Multi-Messenger Astronomy and Astrophysics \hspace*{65pt} \linebreak
  
\textbf{Principal Author:}

Name:	Sally Oey
 \linebreak						
Institution:  University of Michigan
 \linebreak
Email: msoey@umich.edu
 \linebreak
Phone:  734-936-7885
 \linebreak
 
\textbf{Co-authors:} Tom Maccarone (Texas Tech), Fred Walter (Stony Brook), Charles Bailyn (Yale), Jay Gallagher (Wisconsin), Todd Henry (Georgia State), Terry Oswalt (Embry Riddle), Derek Buzasi (Florida Gulf Coast),  J. Allyn Smith (Austin Peay), Rachael Beaton (Princeton and Carnegie Observatories), Jim Webb (FIU), Brad Barlow (High Point), Misty Bentz (Georgia State), Leslie Hebb (Hobart and William Smith), Patrick Kelly (Minnesota), Jedidah Isler (Dartmouth), Michael Meyer (Michigan), John Salzer (Indiana), Simone Scaringi (Texas Tech)
% Claudia Scarlata (Minnesota)
  \linebreak

\textbf{Abstract  (optional):}

Open-access telescopes of all apertures are needed to operate a competitive and efficient national science program.  While larger facilities contribute light-gathering power and angular resolution, smaller ones dominate for field of view, time-resolution, and especially, total available observing time, thereby enabling our entire, diversely-expert community.  
% We argue that smaller telescopes are therefore collectively more productive than larger facilities.  
Smaller aperture telescopes therefore play a critical and indispensable role in advancing science.
Thus, the divestment of NSF support for modest-aperture (1 -- 4 m) public telescopes poses a serious threat to U.S. scientific leadership, which is compounded by the unknown consequences of the shift from observations driven by individual investigators to survey-driven science. 
Given the much higher cost efficiency and dramatic science returns for investments in modest aperture telescopes, it is hard to justify funding only the most expensive facilities.
{\bf We therefore urge the Astro2020 panel to explicitly make the case for modest aperture facilities, and to recommend enhancing this funding stream to support and grow this critical component of the OIR System.}  Further study is urgently needed to prioritize the numerous exciting potential capabilities of smaller facilities,and to establish sustainable, long-term planning for the System.  

Cost category:  
%Ground/Small ($< \$20$M / yr); 
Ground/Medium-Large over 10 years.

\defcitealias{Elmegreen2015}{Elmegreen Report}
\defcitealias{McKee2001}{AANM}
\defcitealias{Blandford2010}{Astro2010}
\defcitealias{Eisenstein2012}{PRC}
\defcitealias{Blandford2006}{2006 NSF-AST Portfolio Review}

\pagebreak

A fundamental sea change is taking place in the enterprise of ground-based, optical and infrared (OIR) astrophysics.  Whereas in the past, observational science was propelled by individual investigators, today, observations are driven ever more by large key projects and surveys \citep[][]{Frogel2010, Abt2017}.  This shift is being driven by advances in both technology and data science.  
It is further accelerated by NSF's withdrawal of support for the individual investigator (``PI science'') mode of allocating observing time.  NOAO has been forced to end general-purpose, public access to most of its own facilities, i.e., the Mayall and Blanco 4-m telescopes and smaller apertures.  These two flagship facilities, together with WIYN, are now dominated by the DESI Survey, Dark Energy Survey, and NN-EXPLORE respectively --- exciting projects, but ones that originate from the priorities set by other agencies, namely DOE and NASA.  These projects have enabled the telescopes to survive and gain vitally important instruments, but at the cost of NSF subordinating some control over its own facilities.
% Allowing other agencies to direct the science conducted on its facilities 
There is a risk that NSF is on a path that compromises its mission to promote basic research through community peer review, a process that is the foundation of our phenomenal success and international leadership in scientific discovery.  And it severely limits community access to observing facilities.

To our knowledge, there has been no community study on the consequences of this shift to survey science at the cost of PI science.  However, several science-driven, strategic plans have stressed the importance of maintaining balanced support for broad community access to modest apertures (see Section~2).  Since circumstances are forcing us in the opposite direction, this may mean risking the optimization of the US science program at a time when astrophysics in Europe and Asia is ascendant.  Unintended consequences pose another serious risk.  In particular, NSF is seriously considering a significant investment in the 30-m class, extremely large telescopes (ELTs).  While public access to these facilities will be an essential boost for the OIR community, the available per capita observing time on them will be extremely low, and mainly dedicated to key projects instead of PI science.  This underscores the fact that public time on the ELTs is not equivalent, nor a substitute for, the public time that NOAO has historically provided.
A comprehensive study is therefore urgently needed to formulate the best strategy for optimizing science productivity in the new, survey-driven era.

\section{Modest-aperture telescopes drive discovery}

Apertures of \textit{all sizes} are needed to operate a productive and efficient national science program.  Telescope discovery space is parameterized by light-gathering power, angular resolution, time resolution, field of view, and available observing time.  The first two are dominated by larger apertures, and the latter three by smaller apertures.  Moreover, a critical driver of discovery is human brain power, which is primarily linked to the last parameter, available observing time.  Maximizing community access is necessary to maximize science.  Combined, these factors cause smaller facilities to dominate raw science productivity.

This is well documented in a variety of sources, including a compendium of discussions from community meetings on small telescopes about two decades ago \citep{Oswalt2003}, including a Lowell Workshop in 1996 and an AAS special session leading up to the 2000 Decadal Survey.  At that time, community concern centered on the closing of NOAO telescopes smaller than 4-m class.  Twenty years later, we are experiencing the same phenomenon with the threshold pushed to the 8-m class.  The dynamic is the same, and arguments and suggestions made in 2003 are still valid and highly recommended reading. 

\begin{figure}
    \centering
    \vspace*{-0.5in}
    \includegraphics[width=13cm]{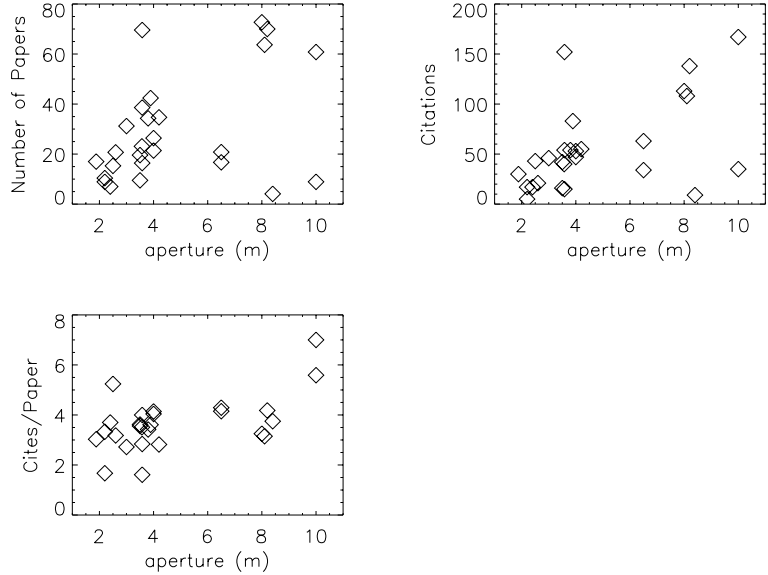}
    \vspace*{-0.5in}
    \caption{Total number of papers, citations, and citations per paper (C/P) for publications in 2008 compiled by \citet{Trimble2010} for general-purpose telescopes. Data for VLT, Keck, Gemini, LBT, and Magellan are divided by the number of telescopes hosted by these facilities. We exclude data where metrics for different aperture telescopes are consolidated for an entire observatory.}
    \label{f_trimble}
\end{figure}

% Since ``large'' class telescopes are defined as the latest generation aperture class, they are always outnumbered by the smaller classes, and so productivity is dominated by the latter.    
Figure~\ref{f_trimble} shows publication metrics for general-purpose telescopes from \citet{Trimble2010}.  We see that for large apertures ($>8$ m), the metrics shown are around twice the values for smaller telescopes, but this relatively modest enhancement by no means causes the large-class facilities to exceed the total productivity of small telescopes.  This confirms trends identified earlier by \citet{Abt2003}.  
\begin{figure}
    \centering
    \includegraphics[width=8cm]{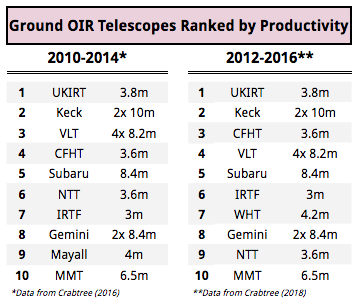}
    \caption{Ranking of general-purpose, OIR telescopes by numbers of papers published during 2010 -- 2014 \citep{Crabtree2016} and 2012 -- 2016 \citep{Crabtree2018}.}
    \label{f_crabtree}
\end{figure}
More recently, \citet{Crabtree2016} ranks telescope facilities by number of papers per telescope published during 2010 -- 2014.  There are 18 OIR facilities in this evaluation, and exactly half of these have apertures above 5 m, and half below. The top 10 ranking is shown in Figure~\ref{f_crabtree}.
% is as follows:  UKIRT, Keck, VLT, CFHT, Subaru, NTT, IRTF, Gemini, Mayall, MMT.  
Again we see the larger and smaller apertures contributing exactly equally to the top 10.  This trend is maintained among the top four.  Moreover, the top-ranked facility is UKIRT, a 3.8-m telescope.  The same trend is seen for the period 2012 -- 2016 \citep[Figure~\ref{f_crabtree};][]{Crabtree2018}.
%Indeed, the three bottom-ranked telescopes out of the 18 are all large-class facilities.  
Thus all apertures contribute equally at all rankings.
This analysis does not include specialized facilities of modest aperture, which are \textit{even more} productive, e.g., SDSS (2.5m) has over 15,000 citations to date on its data release papers, and the 2MASS All-Sky-Release \citep[1.3-m;][]{2MASS} has over 6,000 citations.

Thus, small and mid-aperture telescopes have greater total productivity by these metrics.  {\bf Factoring in their cost-effectiveness dramatically compounds this effect.}  Figure~\ref{f_trimble} suggests that the citation rate is roughly linear with aperture.  \citet{Weaver2003}, therefore finds that, for telescope construction cost scaling roughly as aperture $A^{2.5}$, the dollar cost per citation scales as $A^{1.5}$.  This relation applies whether or not operations costs are included; \textit{science on an 8-m costs almost 3$\times$ as much as on a 4-m, and 8$\times$ as much on a 2-m}.  Thus, while proponents of large telescopes argue that these facilities generate more important science, the citation rates suggest that nevertheless, we see a steep diminishing return on investment with larger apertures.  

Simply put, modest aperture telescopes collectively have been more productive and cost-effective than large-class facilities.  There is an unfortunate misperception that ``small telescopes produce small science."  Yet, some of the most important discoveries of our time were made with modest apertures:  the first exoplanet, 51 Pegasi \citep[Haute Provence 1.9-m;][]{Mayor1995}; MACHOs as a 20\% component of dark matter \citep[Mt. Stromlo 1.3-m;][]{Alcock2000}; and the existence of dark energy \citep[CTIO 1.5-m and 4-m, among others;][]{Riess1998}.  
% Most of these telescopes were considered small even at the time of these revolutionary discoveries in the 1990s.  
More recently, the electromagnetic counterpart of the first gravitational wave detection was initially identified with a 0.83-m telescope at CTIO \citep{Diaz2017}.
% as listed by \citet{Najita2019} in their Astro2020 white paper.
We need only look to current exoplanet searches, supernova transient projects, asteroseismology studies, and solar system planetesimal surveys to see the on-going, essential contribution of smaller telescopes.  In their Astro2020 white paper, \citet{Najita2019} compare the respective productivity of small and large telescopes as ``growth stocks'' versus ``value stocks.''  This is an apt analogy, underscoring the fact that smaller telescopes are the ones that truly drive the entire scientific enterprise, while large facilities provide fundamental quantum advances in opening critical discovery space.  Another relevant analogy is to note that desktop computers are needed as much as high-performance computing; different scale technologies simply play different but complementary roles. Both need support.

\section{The Ground-Based OIR System}

Conceptualizing the public, ground-based, OIR telescopes as ``The System'' grew from before the 2000 decadal review \citep[AANM;][]{McKee2001} and was established by a series of System Workshops that took place between \citetalias{McKee2001} and the 2006 NSF Senior Review \citep{Blandford2006}.  
Based on comprehensive community consultation, the Senior Review strongly endorsed support for modest-size telescopes: {\it ``Instead of further divestiture of the CTIO and KPNO telescopes,... [the Senior Review] recommends that NOAO should acknowledge a responsibility to lead the utilization of small and mid-sized telescopes in the public sector to address the new and emerging opportunities...''}  
This guidance was based on three arguments:  1.~The synergy between large and small aperture facilities, anticipating an \underline{increased} role for smaller facilities due to, e.g., time domain science; 2.  More than half of US astronomers and their students do not have access to private facilities; 3.  Smaller telescopes provide critical test beds for new technology. 
Recommendation 2 of the Senior Review was:
{\it ``The OIR Base program should be led by NOAO. It should deliver community access to an optimized suite of high-performance telescopes of all apertures through Gemini time allocation, management of TSIP and operation of existing or possibly new telescopes at CTIO in the south and KPNO or elsewhere in the north.''}

Pursuant to the Senior Review, NOAO constituted two committees,
Renewing Small Telescopes for Astronomical Research (ReSTAR), and Access to Large Telescopes for Astronomical Instruction and Research (ALTAIR) to recommend strategic plans for modest and large facilities, respectively.
ReSTAR emphatically stressed the need for comprehensive public access to small and mid-aperture telescopes, again emphasizing that modest- and large-scale facilities play fully complementary roles in astrophysical discovery.  
The report echoed findings from \citetalias{McKee2001} that the new era of survey science would increase the demand for modest apertures, and that the available public access was many times less than contemporary community requires.  
ReSTAR thus envisioned the System as a network of telescopes coordinated by NOAO, recommending a {\it doubling} in the number of public 2 -- 4-m telescopes \citep{Pilachowski2008}.
ReSTAR also recommended extending the Telescope System Instrumentation Program (TSIP) to mid-scale telescopes.  
This program, originally proposed by \citetalias{McKee2001}, did become a backbone of the System, providing substantial amounts of public access to a variety of facilities until the program was discontinued after 2012.

The 2010 decadal review \citep[Astro2010;][]{Blandford2010} acknowledged community open access to telescopes ``with apertures up to 8 meters'' as a ``critical need.''  However, one of their Conclusions was that NSF's funding allocations needed to be rebalanced between NOAO's telescopes, existing partnerships, and future facilities, such that the last should benefit.  The report also suggested that NOAO's mid-aperture telescopes would become less relevant in the ELT era, an argument that the report did not substantiate and with which we disagree \citep[e.g.,][]{Pilachowski2008}.  This led Astro2010 to suggest that NSF reevaluate NOAO's role in the community.

Revised budget realities led to the 2012 NSF-AST Portfolio Review (PRC), which seized upon these sentiments and led to the recent NSF divestment from public access to the KPNO 2.1-m, WIYN, and Mayall telescopes \citep[Recommendation 10.6;][]{Eisenstein2012}.  The panel was charged with an extraordinarily difficult task of making hard choices to slash expenditures and programs; however, there was little quantitative discussion of comprehensive productivity metrics in the context of the multi-aperture OIR System.  In particular, there was little discussion about balancing current productivity against future needs, and there was an explicit shift such that the recommendations {\it ``reduce the {\underline{quantity}} of time available via open access to moderate/small OIR telescopes, but the portfolio aims to increase the value of each night, by raising the average aperture, site quality, and instrumentation quality of AST-funded OIR observations...''}.  The \citetalias{Eisenstein2012} therefore made the decision to disenfranchise half of the astronomical community who only have access to telescope facilities via the public system.  NOAO's System Roadmap Committee issued a strong Statement \citep{Soifer2012} expressing concern about these issues.

With NSF's assets significantly reduced, the NRC commissioned a new study to develop a repositioning strategy for the OIR System led by Debra Elmegreen \citep[``Elmegreen Report";][]{Elmegreen2015}.  The panel carried out a new, quantitative inventory of the public OIR System in the context of the historical strategic plans.  They revalidated the fundamental need to support all apertures.  Their \underline{Recommendation 1} attempted to recover public access by charging NOAO to oversee a program for observatories to exchange telescope time and instruments, while managing a central TAC process.  \underline{Recommendation 2} was for NOAO to constitute a System organizing committee to maintain an ongoing, community-wide planning process.  
At the time of the \citetalias{Eisenstein2012} in 2012, NOAO offered on the order of 600 nights per semester on 2 -- 6-m telescopes \citep{Soifer2012}.  The 2019B Call for Proposals offers only 131 nights in this aperture class, a drastic reduction.  An additional 135 nights are offered on 8-m class telescopes, and $\sim$200 nights are available on 1 -- 2-m telescopes via SMARTS and Las Cumbres Observatory.

\section{The Ground-based OIR Future}

Recommendations 1 and 2 from the \citetalias{Elmegreen2015} remain to be effectively implemented.  As stressed above in Section~1, by far the most cost efficient way to maximize scientific productivity is to invest in publicly accessible, modest aperture telescopes.  
As outlined above, this is driven by several factors:  1. Large telescopes alone are greatly restricted in multi-parameter discovery space; 2. Public access enables the entire community.  3. Modest facilities and their instrumentation are 1 -- 2 orders of magnitude cheaper to build and operate than large ones.  Training of students 
\citep[e.g., white paper by ][]{Whelan2019}
and technical personnel, as well as hosting technology trials are additional critical roles with broad, long-term consequences that are extensively called out in previous studies.

These points have been demonstrated and stressed in several strategic planning papers over the last 20 years as documented in Section 2, based on a variety of studies and comprehensive community input.  We simply cannot afford to ignore the critical role of modest-aperture facilities if we are to maintain the health of our field and remain internationally competitive.

{\bf We therefore urge the Astro2020 panel to explicitly make the case for modest aperture facilities, and to recommend an enhancement of this funding stream to support and grow this critical component of the OIR System.}  As stressed above, a small budget, perhaps on the order of \$10M / year, would yield outstanding return on investment.
The immediate goal should be to more aggressively carry out recommendations from the \citetalias{Elmegreen2015}, especially Recommendations 1 and 2, to expand public access to modest apertures as rapidly as possible. 
The OIR System also needs sustainable, long-term strategic plans, along with guidance to NSF for different budget scenarios.  

% DLB note: where does this number come from? \cite{VanBelle2004} would predict construction costs as about \$2M for a 2 meter and \$10M for a 4 meter, and \cite{Abt2012} operating costs of about $\$2 - 6$M annually for each of these, though those operating costs seem high to me. In any case, we should probably provide some context for the \$10M number. }
% This amount is crudely on the order of support for the NOAO telescopes that have been divested.  

Further study is needed to evaluate the community's most urgent needs and what the most effective solutions are, now that public time on mid-size apertures is relatively unavailable.  Based on a rough estimate of \$180k~yr$^{-1}\ \rm m^{-2}$ for operations costs on existing modest-aperture facilities, \$10M / yr could support a total of two 4-m, four 2-m, and eight 1-m telescopes, which could include allowance for instrument and facilities development and student training programs. This calculation is merely illustrative;
in the Appendix, we list a variety of needed capabilities in the context of the currently available System.  
It will be important to study the balance between long-term and smaller projects, as there will be enormous demand for both in the wake of facility-scale surveys.
Which capabilities are deemed essential should be determined deliberately in close consultation with the community of users.  

Cutting-edge technology on smaller facilities is often a fraction of the cost on larger telescopes, but can dramatically improve system performance, thus effectively increasing the aperture.  Upgrading detector and throughput efficiency yield a drastic increase in signal-to-noise.  Image improvement technology has a similar effect, perhaps including adaptive optics (AO) on mid-size telescopes \citep[e.g., Robo-AO, now on UH 2.2-m;][]{Riddle2015}.  Improvements in observing efficiency likewise can yield substantial gains in observing time, for example, rapid slew capability and more efficient target and guide star acquisition.  
Furthermore, as read noise drops below sky background levels, imaging data from arrays of small telescopes can be combined without loss of sensitivity \citep[e.g., as in the BlackGEM project;][]{Bloemen2015}, and photonic linking devices are beginning to enable the same for spectroscopy \citep[see white paper by][]{Eikenberry2019}.  Such facilities can be designed with operational flexibility so that, for example, a telescope array with a total aperture equivalent to a 4-m could also function as 16 individual 1-m telescopes.

% {\color{blue}
Forming university consortia to assume operations of existing telescopes is a proven model for preserving existing facilities.  Successful examples include the Small and Moderate Aperture Research Telescope System (SMARTS) at CTIO, and the Southeastern Association for Research in Astronomy (SARA).  The latter assumed operations of a 1-m telescope at KPNO beginning in 1995 and refurbished it to be remotely accessible \citep{Keel2017}.  The healthy, 15-institution consortium now operates three remote-access telescopes at KPNO, CTIO and La Palma that serve about 200 users per year at an operational cost of less than \$200 per night.  
New facilities can also be built in partnership with the national observatories, as in the case of the 3.5-m WIYN at KPNO.  SMARTS, in operation since 2003, is supported more by a mix of individual PIs, institutions, and NASA funds.  The consortium is being forced to scale back operations by closing public access to two (1.0-m and 1.3-m) of their four telescopes.
Anecdotal reports indicate that NSF grant review panels are increasingly reluctant to fund support for small telescopes, which makes cost-sharing models difficult for small institutions.  Travel support for observing is also a perennial issue for investigators at smaller institutions, and could be alleviated by considering institutional or group grants.  Alternatively, incentivizing and supporting queue, service, and robotic observing may prove feasible in some models.  Consortium sustainability depends on a variety of factors, including operations budget, funding sources, number of nights per member, existence of long-term projects, and availability of contingency funds.  
A review of funding should examine all the relevant factors, including the role of PI funding for observing support.  
% }

As stressed in Section 2, small-scale funding on modest-aperture facilities yields dramatic science returns.  It is therefore hard to justify funding only the most expensive facilities.
NOAO has already made good efforts to develop time exchanges.  Limited public access is currently available this semester (2019B) on Subaru and AAT (3 and 5 nights, respectively).  LBT is offering 8 nights through the discontinued TSIP program.  Through MSIP, Las Cumbres and the CHARA interferometer are offering on the order of 144 and 30 nights, respectively. 
% NOAO is apparently still a consortium partner in SMARTS
These efforts are commendable, but there is a great need to grow access, especially in the 2 -- 4-m class.  For example, NOAO (soon NCOA\footnote{NOAO is in the process of transitioning under the National Center for OIR Astronomy}) could help coordinate and solicit submissions to the MSIP program to ensure that modest apertures continue to participate.  The enormous success of TSIP was due to the two-fold benefit of updating instrumentation on smaller telescopes, which is critical for their competitiveness; and providing community access in exchange for funding.  More focused advocacy and coordination could generate more buy-in from private observatories, motivating them to upgrade via MSIP.

System coordination could also provide support for observatories to improve their operations and greatly promote observing efficiency.  For example, this could include 
development of standard observing templates for a variety of applications, as pioneered at, e.g., UKIRT and YALO (the predecessor to SMARTS).  These enable service observing by less experienced observers, alleviating the need for hiring dedicated service observers.  Some observatories also may have the means, but lack the FTE to develop queue observing systems, and could use System coordinated help.  Even fairly basic needs like installing software to identify guide stars can be facilitated and encouraged.  If appropriate, certain strategies and technologies may be prioritized or incentivized through system coordination. 
For example, telescopes equipped with only a single instrument drastically reduce operations and maintenance costs, and lend themselves to remote and/or robotic operation.  
As small telescope arrays become more established, it may even be desirable to consider working toward standard telescope control systems, which would greatly facilitate observing efficiencies by users and implementation of robotic systems and remote observing.

Encouragingly, the creation of NCOA and its organizational structure appropriately positions Mid-Scale Observatories as an equal partner to LSST, Gemini, and the Community Science and Data Center.  This will enable the much-needed, dedicated attention and advocacy for this System component.  {\bf We strongly urge that the Astro2020 report explicitly stress the critical need to commit the modest funding line needed to grow support for the smaller-scale facilities over the next decade.}  These have supported the bulk of the community, and are therefore the heart of PI-driven science.
It is sobering to note that in some cases (e.g., SMARTS), scientists are supporting facilities by writing personal checks.
We also recommend a community study to prioritize funding needs for modest aperture telescopes and develop a sustainable, long-term strategic plan for the OIR System. Such a study should include clarifying the required level of investment relative to other NSF priorities.
The \citetalias{Elmegreen2015} Recommendation 2 to provide on-going, science-driven planning could then be established from this starting point.

To paraphrase the 2006 Senior Review \citep{Blandford2006}, {\it innovative ideas, advanced technology and clever data handling can be more important than simply increasing collecting area}.
Like the mass functions of stars and galaxies, although the smaller objects are less luminous, they dominate the mass of the entire system.  We ignore them at our peril. 

% 			Suggested final sentence: 
% Through these and other ways, we urge the Astro2020 discussions to include creative ways of re-establishing and  the spirit of AURA's original mission to provide access to astronomical facilities to users that cannot afford their own.  
% \bigskip\bigskip\bigskip

\vspace*{0.7in}
\pagebreak

\appendix{{\bf\Large Appendix:  Specific Needs for Observational Capabilities}}

Adequately supporting modest aperture telescopes requires maintaining and upgrading their technology and instrumentation.  {\it Modest-aperture telescopes will be in enormous demand for follow-up observations in both survey mode and smaller projects.}  The ReSTAR Report therefore recommended a range of capabilities on small and mid-aperture telescopes in both hemispheres \citep{Pilachowski2008}, especially low and high-resolution spectroscopy, and optical and IR imaging.  Here we outline some anticipated needs for observational capabilities based on typical science drivers, relative to the current ground-based, public OIR System.  This description is meant to be more illustrative than exhaustive.  Future study is needed to explore and prioritize a more complete range of capabilities and determine a recommended funding level relative to other NSF priorities.

{\it Imaging:}
In the era of LSST, ZTF and PanSTARRS, the role of small-to-medium telescopes in imaging projects will evolve.  
Single-epoch projects may be dominated by emission-line imaging; and by infrared surveys deeper than provided by 2MASS, and in regions not covered by the initial set of VISTA surveys.  Additionally, there will be demand for imaging at high angular resolution, both for fainter objects that require AO, and for bright objects that require interferometry.
At the present time, the national System has a good set of wide-field, optical imagers on a broad range of telescope apertures, but the filter set is insufficient. There is also no wide-field ($>20^\prime\times20^\prime$) infrared imaging and relatively little high time-resolution instrumentation, nor instruments that can simultaneously observe in multiple filters.  The System currently has reasonably good capabilities for rapid response on small telescopes through Las Cumbres, but lacks the fastest slewing, robotic telescopes that are needed for prompt follow-up of transients.

{\it Spectroscopy:}
While SDSS-V will dramatically expand the database of spectra, LSST will create a new need for mass-scale spectroscopy.  
% While many LSST-generated targets will be too faint for mid-sized telescopes, the b
Brighter targets generated by LSST will be spread over the sky in a manner that is not conducive to multi-object spectroscopy, meaning that many smaller telescopes will be needed to follow them up.  Time-domain spectroscopy is also needed for targets of opportunity triggered from other wavelength ranges.  Furthermore, a broad range of parameter space, including sky coverage, will be left uncovered.  For example, no mass-scale infrared spectroscopic surveys are currently planned. % {\color{blue} 
Even today, the current System would be greatly enhanced by offering queue-scheduled, moderate-resolution spectroscopy on a 2 -- 4-m class telescope for, e.g., supernova and GRB follow-ups and AGN reverberation mapping.  Modernizing spectrograph efficiencies would be especially productive, as demonstrated by the performance of the CHIRON echelle on the SMARTS 1.5-m and
Goodman spectrograph on SOAR.

{\it Time-domain:}
Time-domain astronomy is generally less compatible with classical observing.
We note that NOAO's oversubscription rates have been higher on the queue scheduled telescopes and 
% more recently on the 
robotic telescopes than for classical observing.  This 
% may be, in part, because of a desire on the part of many PIs to avoid the time and expense of observing trips, and very short programs that simply cannot be done in classical mode.  
is at least partly because these telescopes enable monitoring campaigns.  As time domain astronomy grows in scope, these types of programs are likely to become more important. 
LSST will be limited in cadence, filters, and northern sky coverage.  For example, infrared monitoring applications (e.g., variable sources in star forming regions) will remain. 
Additionally, there will be need for high-cadence monitoring (e.g., searches for orbital periods of short-period binaries); and long, uninterrupted monitoring to sample variability timescales (e.g., accretion variability and white dwarf pulsations).  As an example,
the Whole Earth Blazar Telescope, a global network of small telescopes, achieved a continuous 72-hour light curve of a highly variable blazar, 0716+71, with 1-minute time resolution \citep{Bhatta2013}. 
Long-term, multi-year monitoring programs greatly benefit from access to consistent instrumentation, as shown by, e.g., studies of the nearest red, brown and white dwarfs \citep{Henry2018}, novae \citep{Walter2012}, and X-ray binaries \citep{Buxton2012}.
Such applications are well suited to small- and medium-sized telescopes, and cannot be done by LSST.  
% Moreover, many of these cannot be done at all on large telescopes because of, e.g, bright saturation limits or slow slew speeds; and/or they represent poor uses of time on larger telescopes.
The availability of substantial amounts of time on the Las Cumbres network is an excellent step;  however, public access will expire in 2020 and it represents only a fraction of what will be needed in the LSST era.

Rapid response to targets of opportunity (TOOs) requires a broad range of instrument capabilities available at all times (e.g., OIR counterparts of gamma-ray bursts).  
This can be done by having many telescopes running in parallel, each with different instruments; or with individual telescopes having either versatile, multi-mode instruments or rapid instrument changing capability, as can be done with instruments mounted at Nasmyth ports.  It is also important to have the ability to schedule TOO programs on very short notice.  NOAO's ANTARES event broker and AEON observing system for LSST transient follow-ups are welcome examples of this essential infrastructure development.  We note that the robust TOO program of NASA's {\sl Swift} mission, which is predominantly scheduled ``on the fly" rather than by panels, is widely regarded as one of its core strengths.

{\it Other capabilities:}
Other observing modes are also worth considering, such as polarimetry and speckle imaging.  Polarimetry, in particular, represents a relatively unexploited frontier for electromagnetic information and hence has excellent growth potential.  It is perhaps well suited to mid-size aperture telescopes, since the required photon fluxes are fairly high, while demand is currently low on larger facilities. However, polarimetry can done on smaller apertures for a variety of science applications, as demonstrated by \citet[][on a 0.35-m!]{Bailey2017} and \citet[][on a robotic 1.5-m]{Ramaprakash2019}.  
Furthermore, there are important scientific areas in which small telescopes can outperform  large ones.  For example, slew speeds for small telescopes can be dramatically faster.  This is important for fast acquisition of transient phenomena like gamma-ray bursts, shock breakouts, and gravitational wave merger events; and also for efficiency in observing large samples of brighter objects.  In addition, many classes of variability studies (e.g., light curves of accreting objects, or stellar pulsations of stars) require time resolution on the order of seconds or shorter, which can be difficult or impossible for brighter objects on large telescopes.  A variety of other observing capabilities should also be examined.

% Smaller telescopes also offer the ability to do photometric monitoring of very bright stars without defocusing. !!!

\pagebreak
% \textbf{References}
\small
\bibliographystyle{apj}
\bibliography{whitepaper.bib}

\end{document}